\documentclass[11pt,a4paper]{article}
\usepackage[german,english]{babel}
\usepackage{epsfig}
\usepackage{epic}
\usepackage{eepic}
\usepackage{latexsym}
\usepackage{amssymb}
\usepackage{amsmath}
\usepackage{vmargin}
\usepackage{amsthm}
\def\refeq#1{(\ref{#1})}

\def\Tr{{\rm Tr}}
\def\KK{\varkappa}
\def\lax{x}
\def\Y{Y}
\def\p0{\nu}
\def\i{{\rm i}}

\newcommand{\erw}[1]{\left<#1\right>}

\newcommand{\abs}[1]{\left|#1\right|}
\newcommand{\delzwei}[2]{\frac{\partial^{2}#1}{\partial#2^{2}}}       
\newcommand{\rundk}[1]{\left( #1 \right)}

\newcommand{\gesk}[1]{\left\{ #1 \right\}}
\newcommand{\ket}[1]{\left|#1\right>}

\newcommand{\delieins}[1]{\frac{\partial}{\partial #1}}

\newcommand{\ska}[3]{\left<#1|#2|#3\right>}

\def\id{{\text{1} \kern-.26em \text{l}}}

\newcommand{\ham}{\mathcal{H}}

\def\e{{\rm e}}
\def\be{\begin{equation}}
\def\ee{\end{equation}}
\def\bea {\begin{eqnarray}}
\def\eea {\end{eqnarray}}

\begin{document}
\title{On the finite temperature Drude weight of the anisotropic Heisenberg chain} 

\author{J. Benz, 
T. Fukui\thanks{Department of Mathematical Sciences,
Ibaraki University, Mito 310-8512, Japan},
A. Kl\"umper\thanks{Theoretical Physics, Wuppertal University, Gauss-Strasse 20,
D-42097 Wuppertal, Germany}, 
C. Scheeren}


\maketitle
\begin{abstract}
We present a study of the Drude weight $D(T)$ of the spin-1/2 $XXZ$ chain in
the gapless regime. The thermodynamic Bethe
ansatz (TBA) is applied in two different ways. In the first application we
employ the particle basis of magnons and their bound states. In this case we
rederive and considerably extend earlier work in the literature. However, in the course
of our investigation we find arguments that cast doubt on the applicability
of the TBA in this case. In a second application by use of the spinon and
anti-spinon particle basis we obtain completely different results.  Only for
anisotropy parameter $\Delta$ close to 0 we find that $D(T)$ is a monotonously
decaying function of temperature. For $\Delta$ close to 1 the behaviour is
entirely different showing a finite temperature maximum. Also for
the isotropic antiferromagnetic chain ($\Delta=1$) the results for $D(T)$ are
finite for $T=0$ as well as for $T>0$ with an infinite positive slope at $T=0$.
\end{abstract}

\section{Introduction}
In this work we are studying transport properties of the spin-1/2 Heisenberg
chain with longitudinal anisotropy ($XXZ$ chain). Depending on the
representation of the system the quantity of interest is the spin conductivity
of a quantum spin system with anisotropic spin exchange or the electrical
conductivity of a system of spinless fermions with density-density
interaction.  Both representations are related by a Jordan-Wigner
transformation. The recent interest in these quantities has several
reasons. We want to mention only two of these. First, in low-dimensional
systems the question has been raised whether spin diffusion exists or not and
the role of integrability for anomalous transport properties was
discussed. (For a review see \cite{ZP03}.)  Second, the Drude peak at zero
frequency in the dynamical conductivity is in principle accessible to
analytical studies.\\

Let us consider a Hamiltonian of the form
\begin{equation}
\hat{H}=\hat{H}_0+\hat{V}=-t\sum_j\left(\e^{ieA_x(j,t)}c_{j+1}^\dag c_j + 
\e^{-ieA_x(j,t)}c_j^\dag c_{j+1}\right)+\ham_{int}
\label{HamilA}
\end{equation}
representing a one-dimensional system of length $L$ with periodic
boundary conditions subject to a vector potential
$A$. In our case $\ham_{int}$ describes density-density interactions and
hence does not depend on the vector potential.

Within Kubo theory \cite{1}, i.e. linear response for \refeq{HamilA} in $A$, 
the dynamical conductivity is obtained in
terms of current-current correlation functions.  The Drude weight $D$
is the zero frequency distribution of the dynamical conductivity
$\sigma=D\delta(\omega)+\sigma_{\mbox{\tiny reg}}$ and has a spectral
representation in terms of eigenstates and eigenvalues of the 
system with vector potential $A=0$
\be
D=\frac{1}{2L}\gesk{\erw{-\hat{T}}- 2\sum\limits_{m\not = n} p_n
\frac{\abs{\ska{n}{\hat{j}}{m}}^2}{\epsilon_m-\epsilon_n}},
\label{spectralD}
\ee 
where $\hat{j}$ is the current operator, $\erw{\hat{T}}$ is the thermal
expectation value of the kinetic energy, and $p_n=\e^{-\epsilon_n/T}/Z$
is the Boltzmann weight for the eigenstate $\ket{n}$ of the
Hamiltonian with energy $\epsilon_n$. \\ 

On the other hand, a static magnetic flux leads to a site and time independent
vector potential $A_x(j,t)=\phi/e$ with characteristic dependence
of the eigenvalues on $\phi$. Denoting the Hamiltonian \refeq{HamilA}
by $\hat{H}(\phi)$ we find in second order pertubation theory in $\phi$
\be
\epsilon_n(\phi)=\ska{n}{\hat{H}(0)}{n}-\phi\ska{n}{\hat{j}}{n}
-\phi^2
\sum\limits_{m\not =n} \frac{\abs{\ska{n}{\hat{j}}{m}}^2}{\epsilon_m-\epsilon_n}-\frac12 \phi^2 \ska{n}{\hat{T}}{n}.
\ee
Comparing second order terms we see that 
\begin{equation}
D=\frac{1}{L}\sum_n p_n\frac12\delzwei{\epsilon_n(\phi)}{\phi}\Big\vert_{\phi\to 0}.\label{curv}
\end{equation}
This is the generalization of Kohn's result \cite{2}, see also
\cite{ShasS90,Zvyagin} to finite temperature \cite{CastZoPr95}.  In this way
the Drude weight $D$ is connected to the twist $\phi$ in the boundary
conditions caused by the applied external field.\\

The expression \refeq{curv} is very interesting as it allows for the
calculation of the Drude weight just from the eigenvalues of the
Hamiltonian without the knowledge of matrix elements. This is
essential for analytic calculations of integrable systems rendering
the task feasable although the remaining work is still formidable.
In \cite{FujKa98} an extension of the traditional Thermodynamic Bethe Ansatz
(TBA) to cover the mean curvature of energy levels was presented. This
method was applied to the spin-1/2 $XXZ$ chain in \cite{Zotos98}. The 
numerical evaluation of the Drude weight showed a curious behaviour
especially for the isotropic chain suggesting that $D(T>0)=0$.\\

Our own interest in the finite temperature Drude weight was 
triggered by the peculiar findings of \cite{Zotos98}. The treatment of the
(strictly) isotropic Heisenberg chain within the TBA approach is rather
challenging as it involves infinitely many functions to be solved from
infinitely many non-linear integral equations. Hence, we suspected that the
findings in \cite{Zotos98} might be based on inappropriate numerical
treatments of these integral equations. This however, is not the case as will
become clear in section 3.\\

In Sec.2 we present some necessary elements of TBA and the generalization for
calculating the Drude weight along reference \cite{FujKa98}. In Sec.3 we
describe the application of the TBA approach to the Heisenberg chain on the
particle basis of magnons and their bound states (strings) somewhat
following the treatment of \cite{Zotos98}. Our main technical achievement in
this section is that we manage to reduce the many integral equations to just
two. Among other things, this allows for much simplified numerical
calculations in comparison to \cite{Zotos98}, however confirming the obtained
data and the steep drop of the Drude weight as discussed above.  In section 4,
we apply the extended TBA method on the basis of spinons and
anti-spinons. The resulting equations are studied analytically and numerically 
yielding results that are completely different from those of section 3 and
ref. \cite{Zotos98}. Most of the discussion of our results is presented in section
5. Some more technical material related to the derivation presented in section
4 can be found in the appendix.


\section{TBA formalism for free energy and Drude weight}

We consider a system of particles with {\it bare} energy
$\epsilon_\alpha(\lax)$ and momentum $p_\alpha(\lax)$
parametrized by the spectral parameter $\lax$. The index $\alpha$ 
labels the species of particles. As we do not use in this work the notion of the
{\it dressed} energy function we take the liberty of dropping
the commonly used upper index $0$ ($\epsilon_\alpha(\lax)$ instead of
$\epsilon^{(0)}_\alpha(\lax)$).

Furthermore we have diagonal scattering of two particles
of species $\alpha$ and $\beta$ with scattering phase $\Theta _{\alpha
\beta}(\lax_\alpha-\lax_\beta)$. The quantization condition for 
an eigenstate characterized by a set of particles with rapidities 
$\lax_{\alpha k}$ for twisted boundary condition with angle $\phi$ reads
\begin{eqnarray}
Lp_\alpha(\lax_{\alpha k}) + \sum_{\beta l} 
\Theta_{\alpha\beta}(\lax_{\alpha k}-\lax_{\beta l})=
2\pi I_{\alpha k}+\phi_\alpha,
\label{quant}
\end{eqnarray}
where $\phi_\alpha$ is a multiple of the applied twist angle
$\phi_\alpha=\phi\cdot n_\alpha$ with some (integer) number $n_\alpha$. (Note
that Bethe ansatz equations usually take such a form.)
We introduce the counting function $Z_\alpha$
\begin{equation}
Z_\alpha (\lax) := \frac{1}{2\pi } p_\alpha(\lax) + 
\frac{1}{2\pi L}\sum_\beta  \sum_l  \Theta _{\alpha \beta } 
(\lax - \lax _{\beta l})-\frac{\phi_\alpha}{2\pi L}. 
\end{equation}
The set of solutions to \refeq{quant} for $\lax_{\alpha k}$ in an
interval of width $\Delta\lax_\alpha$ comprises actual rapidities as well
as holes with density functions $\rho_\alpha$ and $\rho_\alpha^h$ the sum
of which is related to the counting function $Z_\alpha$
\begin{equation}
\#{\rm particles}+\#{\rm holes}=(\rho_\alpha+\rho_\alpha^h)
\Delta\lax=Z_\alpha(\lax+\Delta\lax)
-Z_\alpha(\lax),
\end{equation}
or explicitly
\begin{equation}
\rho _\alpha (\lax )+\rho _\alpha ^h(\lax) =  \frac{1}{2\pi } p'_\alpha(\lax) + \frac{1}{2\pi L}\sum_\beta  \sum_l  \Theta _{\alpha \beta }' (\lax - \lax _{\beta l}). 
\end{equation}
The summation over rapidities on the r.h.s. can be written in the 
thermodynamic limit in terms of integrals over $\rho_\beta$
\begin{equation}
(1+\eta_\alpha)\rho_\alpha (\lax)=
\frac{1}{2\pi } p'_\alpha(\lax) + \frac{1}{2\pi}\sum_\beta 
\KK_{\alpha \beta }\ast \rho _\beta, \quad
\KK_{\alpha \beta }:=\Theta_{\alpha \beta }',\quad
\eta_\alpha:=\frac{\rho^h_\alpha}{\rho_\alpha}.
\label{rho1}
\end{equation}
This set of integral equations is equivalent to the Bethe ansatz equations
\refeq{quant}. The state representing the macrostate for finite temperature $T$
is obtained from the minimization of the free energy functional 
\begin{eqnarray}
f&=&e-Ts\nonumber\\
&=&\sum_\alpha\Big(\int_{-\infty}^{\infty}\epsilon_\alpha
\rho _\alpha d\lax
-T\int_{-\infty}^{\infty}\left([\rho _\alpha+\rho _\alpha ^h]\ln(\rho _\alpha +\rho _\alpha ^h)-\rho _\alpha\ln\rho_\alpha-\rho_\alpha^h\ln\rho _\alpha ^h\right)d\lax\Big).\nonumber\\
\label{fen}
\end{eqnarray}

This results into the non-linear integral equations ({\it thermodynamic Bethe
ansatz equations})
\begin{equation}
\ln \eta_\alpha(\lax
)=\beta\epsilon_\alpha
(\lax)-\frac{1}{2\pi} \sum_\beta \KK_{\alpha \beta}\ast \ln\rundk{
1+{\eta_\beta}^{-1}},
\label{rho2}
\end{equation}
where we have used the symmetry property $\KK_{\alpha \beta}(x):= \KK_{\beta
\alpha}(-x)$.  By use of this set of equations we can simplify the expression
for the free energy \refeq{fen} yielding
\begin{equation} 
-\beta f=\frac{1}{2\pi}\sum_{\alpha}\int_{-\infty}^{\infty}
p_\alpha'(x)\ln(1+\eta^{-1}_\alpha(x))\,\,dx.
\label{FreeEn0}
\end{equation}
The equations \refeq{fen}-\refeq{FreeEn0} are valid for finite magnetic field
$h$ if the term $\epsilon_\alpha$ is replaced by $\epsilon_\alpha-h n_\alpha$
where $n_\alpha$ is the same number that occurred below \refeq{quant}.
We do not give the details of these calculations. The interested reader
is referred to the book \cite{Takahashi99}. 

Here we are more concerned with the finite size analysis of the rapidities for
which we closely follow the treatment of \cite{FujKa98}. For $\lax_{\alpha
j}$ we make the ansatz
\begin{equation}
\lax_{\alpha j}=\lax_{\alpha j}^\infty+\frac{g_{\alpha
j}^{(1)}}{L}+\frac{g_{\alpha j}^{(2)}}{L^2} =\lax_{\alpha
j}^\infty+\frac{g_{\alpha}^{(1)}(\lax_{\alpha j}^\infty)}{L}+
\frac{g_{\alpha}^{(2)}(\lax_{\alpha j}^\infty)}{L^2},
\end{equation}
with finite size coefficients $g_{\alpha j}^{(1,2)}$ taking the form of
smooth functions $g_{\alpha}^{(1,2)}(\lax)$ in the thermodynamic limit.
Inserting this into the counting function leads to an expansion
\begin{eqnarray}
Z_\alpha (\lax)&=&\frac{1}{2\pi}p_\alpha(\lax)+ \frac{1}{2\pi
L}\sum_\beta \sum_l \Theta _{\alpha \beta } (\lax - \lax
_{\beta l}^\infty)\nonumber\\
&-&\frac{1}{2\pi L}\sum_\beta \sum_l
\Theta _{\alpha \beta }' (\lax - \lax _{\beta
l}^\infty)\frac{g_{\beta l}^{(1)}}{L}-\frac{\phi_\alpha}{2\pi
L}\nonumber\\
&+& \frac{1}{2\pi L}\sum_\beta \sum_l \Theta _{\alpha
\beta }'' (\lax - \lax _{\beta
l}^\infty)\left(\frac{g_{\beta
l}^{(1)}}{2L}\right)^2\nonumber\\
&-&\frac{1}{2\pi L}\sum_\beta \sum_l
\Theta _{\alpha \beta }' (\lax  - \lax _{\beta
l}^\infty)\frac{g_{\beta l}^{(2)}}{L^2}.
\end{eqnarray}

We identify the $O(1)$, $O(1/L)$, $O(1/L^2)$ contributions to the
counting function
\begin{eqnarray}
Z_\alpha (\lax)&=&Z_\alpha^{\infty} (\lax)+\frac{Z_\alpha^{(1)}}{L} (\lax)+\frac{Z_\alpha^{(2)}}{L^2} (\lax)\\
Z_\alpha^{(\infty)} (\lax)&=& \frac{1}{2\pi } p_\alpha(\lax) + \frac1{2\pi}\sum_\beta \Theta _{\alpha \beta }\ast \rho _\beta (\lax)\\  
Z_\alpha^{(1)}(\lax)&=&-\frac{1}{2\pi }\sum_\beta \Theta _{\alpha \beta }'\ast (g_\beta^{(1)}\rho_\beta)-\frac{\phi_\alpha}{2\pi}\\ 
Z_\alpha^{(2)}(\lax)&=&\frac{1}{2\pi}\sum_\beta \left(\frac12\Theta _{\alpha \beta }''\left(g_\beta^{(1)^2}\rho_\beta\right)-\Theta _{\alpha \beta }'\ast (g_\beta^{(2)}\rho_\beta)\right),
\end{eqnarray}
where we have replaced the summations over $\lax_{\beta l}^\infty$ by
integrals involving the density functions.

For studying the quantization condition \refeq{quant} we have to expand
terms of $Z$ like
\begin{equation}
Z_\alpha(\lax_{\alpha j})=Z_\alpha\left(\lax_{\alpha j}^\infty+\frac{g_{\alpha j}^{(1)}}{L}+\frac{g_{\alpha j}^{(2)}}{L^2}\right).
\end{equation}
leading to
\begin{eqnarray}
Z_\alpha(\lax_{\alpha j})&=&Z_\alpha(\lax_{\alpha j}^\infty)+
\frac1L Z_\alpha'(\lax_{\alpha j}^\infty) g_{\alpha j}^{(1)}\nonumber\\
&+&\frac1{L^2} \left[\frac12 Z_\alpha''(\lax_{\alpha j}^\infty)
g_{\alpha j}^{(1)^2}+Z_\alpha'(\lax_{\alpha j}^\infty)
g_{\alpha j}^{(2)}\right]\nonumber\\
&=&Z_\alpha^{\infty}(\lax_{\alpha j}^\infty)+\frac1L\left[Z_\alpha^{(1)}
(\lax_{\alpha j}^\infty)+Z_\alpha^{\infty'}(\lax_{\alpha j}^\infty)
g_{\alpha j}^{(1)}\right]\nonumber\\
&+&\frac1{L^2}\Big[Z_\alpha^{(2)}(\lax_{\alpha j}^\infty)+ 
Z_\alpha^{(1)'}(\lax_{\alpha j}^\infty)g_{\alpha j}^{(1)}+
\frac12 Z_\alpha^{\infty''}(\lax_{\alpha j}^\infty)g_{\alpha j}^{(1)^2}
\nonumber\\
&+&Z_\alpha^{\infty'}(\lax_{\alpha j}^\infty)g_{\alpha j}^{(2)}\Big]
=\frac{2\pi I_\alpha}{L}\nonumber\\
\end{eqnarray}

This equation imposes for the corrections
\begin{eqnarray}
Z_\alpha^{(1)}(\lax_{\alpha j}^\infty)+Z_\alpha^{\infty'}
(\lax_{\alpha j}^\infty)g_{\alpha j}^{(1)}=0\nonumber\\
Z_\alpha^{(2)}(\lax_{\alpha j}^\infty)+ Z_\alpha^{(1)'}
(\lax_{\alpha j}^\infty)g_{\alpha j}^{(1)}+\frac12 Z_\alpha^{\infty''}
(\lax_{\alpha j}^\infty)g_{\alpha j}^{(1)^2}+Z_\alpha^{\infty'}
(\lax_{\alpha j}^\infty)g_{\alpha j}^{(2)}=0,\nonumber\\
\end{eqnarray}
reading in the thermodynamic limit 
\begin{eqnarray}
(1+\eta_\alpha)\cdot(g_\alpha^{(1)}\rho_\alpha)&=&\frac{\phi_\alpha}{2\pi}+
\frac1{2\pi}\sum_\beta\KK_{\alpha\beta}\ast (g_\beta^{(1)}\rho_\beta)
\nonumber\\
(1+\eta_\alpha)\cdot(g_\alpha^{(2)}\rho_\alpha)&=&l_\alpha'+
\frac1{2\pi}\sum_\beta\KK_{\alpha\beta}\ast(g_\beta^{(2)}\rho_\beta)
\nonumber\\
l_\alpha&=&\frac12(\rho_\alpha+\rho_\alpha^h)g_\alpha^{(1)^2}-
\frac1{4\pi}\sum_\beta\KK_{\alpha\beta}\ast(g_\beta^{(1)^2}\rho_\beta)
\nonumber\\
\end{eqnarray}
Focusing again on the Drude weight we are rather interested 
in the derivatives with respect to $\phi$ which we denote by dots
\begin{eqnarray}
(1+\eta_\alpha)\cdot(\dot
g_\alpha^{(1)}\rho_\alpha)&=&\frac{n_\alpha}{2\pi}+\frac1{2\pi}\sum_\beta
\KK_{\alpha\beta}\ast(\dot g_\beta^{(1)}\rho_\beta),\nonumber\\
(1+\eta_\alpha)\cdot(\ddot g_\alpha^{(2)}\rho_\alpha)&=&\ddot l_\alpha'+
\frac1{2\pi}\sum_\beta\KK_{\alpha\beta}\ast(\ddot
g_\beta^{(2)}\rho_\beta),\\
\ddot{l}_\alpha&=&(\rho_\alpha+\rho_\alpha^h)\dot
g_\alpha^{(1)^2}-\frac1{2\pi}\sum_\beta\KK_{\alpha\beta}\ast(\dot
g_\beta^{(1)^2}\rho_\beta).\nonumber
\label{ddott}
\end{eqnarray}

Next we note the second order term in $\phi$ of the 
energy function
\begin{equation}
E_2=\sum_\alpha\int_{-\infty}^\infty\left(\frac12\epsilon_\alpha{''}
g_\alpha^{(1)^2}\rho_\alpha+\epsilon_\alpha{'}g_\alpha^{(2)}
\rho_\alpha\right)d\lax,
\label{nochkorrektoderI}
\end{equation}
the second derivative of this is   
\begin{equation}
\label{d1}
2D=\ddot E_2=\sum_\alpha\int_{-\infty}^\infty\left(\epsilon_\alpha{''}
\dot g_\alpha^{(1)^2}\rho_\alpha+\epsilon_\alpha{'}\ddot g_\alpha^{(2)}
\rho_\alpha\right)d\lax.
\end{equation}
At this point we want to comment on some potentially fatal problem residing in
the last expressions. By inspection of explicit examples we see that separate
integrals of the individual summands occurring in the integrands of
\refeq{nochkorrektoderI} and \refeq{d1} diverge!

Ignoring this problem we may reformulate the last expression 
\begin{eqnarray}
D&=&\frac12\sum\limits_\alpha\left(
\int_{-\infty}^\infty\epsilon_\alpha{''}(\dot g_\alpha^{(1)^2}
\rho_\alpha)d\lax+\int_{-\infty}^\infty
\epsilon_\alpha{'}\ddot g_\alpha^{(2)}\rho_\alpha d\lax
\right)\nonumber\\
&=&\frac1{2\beta}\sum\limits_\alpha 
\int_{-\infty}^\infty\frac{\dot g_\alpha^{(1)^2}
\rho_\alpha[\frac{\partial}{\partial\lax}\ln\eta_\alpha]^2}
{(1+\eta_\alpha^{-1})}d\lax,
\label{semifinal}
\end{eqnarray}
where in the last line we have used among other things an identity
following from the ``dressed function'' formalism applied to \refeq{rho2}
(after taking the derivative with respect to the the spectral parameter)
and \refeq{ddott}.\\

Expression \refeq{semifinal} was derived in \cite{FujKa98} for the study
of the Hubbard model. Here we like to simplify this expression by showing that
all functions appearing in the integrand of \refeq{semifinal} are simply
related to $\eta_\alpha$.  For this we note the relation
\be
\epsilon_\alpha=J\frac{\sin\gamma}\gamma\,p_\alpha',
\label{bareen}
\ee 
and the validity of \refeq{rho2} for finite field $h$ provided
$\epsilon_\alpha$ is replaced by $\epsilon_\alpha-h n_\alpha$. This yields
\bea
J\frac{\sin\gamma}\gamma\, \rho_\alpha&=&-\frac{1}{2\pi}\frac{\partial}{\partial\beta}
\ln(1+\eta_\alpha^{-1}),\cr 
\dot
g_\alpha^{(1)}\rho_\alpha&=&\pm\frac{1}{2\pi}\frac{\partial}{\partial\beta
h} \ln(1+\eta_\alpha^{-1}),
\label{diffid}
\eea
just because both sides of each equation satisfy the same integral equation.
Putting together \refeq{diffid} and \refeq{semifinal} we find
\be
D=\frac{J\sin\gamma}{4\pi\beta\gamma}\sum\limits_\alpha
\int_{-\infty}^\infty\frac
{(\frac{\partial}{\partial\beta h} \ln\eta_\alpha)^2
(\frac{\partial}{\partial\lax} \ln\eta_\alpha)^2}
{(1+\eta_\alpha)(1+\eta_\alpha^{-1})
\frac{\partial}{\partial\beta}\ln\eta_\alpha}
d\lax.
\label{final}
\ee
This expression is amazingly symmetric as the identity
\be
\frac
{(\frac{\partial}{\partial\beta h} \ln\eta)^2
(\frac{\partial}{\partial\lax} \ln\eta)^2}
{(1+\eta)(1+\eta^{-1})
\frac{\partial}{\partial\beta}\ln\eta}
=-\frac
{(\frac{\partial}{\partial\beta h} \ln\frac 1\eta)^2
(\frac{\partial}{\partial\lax} \ln\frac 1\eta)^2}
{(1+\eta^{-1})(1+\eta)
\frac{\partial}{\partial\beta}\ln\frac 1\eta}
\label{particlehole}
\ee 
may be interpreted (up to the sign change) as invariance under a ``particle
hole transformation'' ($\eta={\rho^h}/{\rho} \leftrightarrow
1/\eta={\rho}/{\rho^h}$). Formulas like \refeq{final} with the additional
minus sign also appeared in \cite{Zotos98}.

\section{TBA and Fusion Hierarchy for the Heisenberg chain}

In general the number of particles entering the TBA formalism as
sketched in the foregoing section is infinite. In the case of the
spin-1/2 Heisenberg chain we have to deal with the single magnon and its 
many bound states (strings). For the special case of 
\be
{\cal H} = J
\sum_{i}[{S}^x_i{S}^x_{i+1}+ {S}^y_i{S}^y_{i+1}+
\cos(\gamma){S}^z_i{S}^z_{i+1}]-h\sum_{i}{S}^z_i
\ee
with anisotropy parameter $\gamma=\pi/\p0$ with integer $\p0$ the number of
bound states is finite and the TBA equations close for a finite set of
$\eta$-functions. This is the case studied in \cite{Zotos98} with the main
equation being \refeq{semifinal} for which the functions $\eta$ etc. are calculated
from \refeq{rho2} etc.\\

We want to calculate the functions appearing in \refeq{final} in a different
manner by employing an alternative approach to the thermodynamics of quantum
systems making use of a lattice path integral formulation
\cite{MSusMIn,KlumTBA}. The quantum system at finite temperature is mapped to
a classical model. In the case of the Heisenberg chain we are led to the study
of a staggered six-vertex model on the square lattice. The size of this
lattice is $L\times N$ where $L$ is the length of the quantum chain and $N$ is
the Trotter number. The partition function of this model is calculated within
a transfer matrix approach.  For these calculations it turns out that the
quantum transfer matrix (QTM), i.e. the column-to-column transfer matrix, is
most appropriate as it shows a spectral gap between the largest and
next-largest eigenvalues even in the limit $L$, $N\to\infty$.

There are mainly two different ways of analysing the spectrum of the QTM: the
Bethe ansatz analysis and the fusion hierarchy. Here we make use of both
approaches. The fusion hierarchy was studied carefully in \cite{KSS98} which we are
going to utilize extensively in the first part of this section. In \cite{KSS98}
a hierarchy of functions $\Y_\alpha(x)$, 
$\alpha=1, 2,...$, is derived from the basic object of the QTM. 
These functions satisfy the functional equations 
\be
\Y_\alpha(x+\i)\Y_\alpha(x-\i)=(1+\Y_{\alpha-1}(x))(1+\Y_{\alpha+1}(x)),
\ee
relating each function $\Y_\alpha$ to two other functions, one with lower
index, the other with higher index.
For anisotropy $\gamma=\pi/\p0$ with integer $\p0$ the functional equations
close at a finite level. (In \cite{KSS98} $\p0$ is denoted by
$p_0$; otherwise we follow their notation closely.) The closure happens as 
$1+\Y_{\p0-1}$ factorizes
\be
1+\Y_{\p0-1}(x)=\left(1+\e^{\beta h \p0/2}K(x)\right)
\left(1+\e^{-\beta h \p0/2}K(x)\right)
\ee
with a function $K(x)$ satisfying
\be
K(x+\i)K(x-\i)=1+\Y_{\p0-2}(x).
\ee
These equations can be put into a more canonical form by the definition
\bea
\eta_\alpha(x)&:=&\Y_\alpha(x), \quad\hbox{for $1\le\alpha\le \p0-2$}\cr
\eta_{\p0-1}(x)&:=&\e^{+\beta h \p0/2}K(x),\cr
\eta_{\p0}(x)&:=&\e^{+\beta h \p0/2}/K(x).
\eea
These $\p0$-many functions satisfy the functional equations
\bea
\eta_\alpha(x+\i)\eta_\alpha(x-\i)&=&(1+\eta_{\alpha-1}(x))(1+\eta_{\alpha+1}(x)),
\quad\hbox{for $1\le\alpha\le \p0-3$}\cr
\eta_{\p0-2}(x+\i)\eta_{\p0-2}(x-\i)&=&(1+\eta_{\p0-3}(x))(1+\eta_{\p0-1}(x))(1+\eta_{\p0}^{-1}(x))\cr
\e^{-\beta h \p0}\eta_{\p0-1}(x+\i)\eta_{\p0-1}(x-\i)&=&
\e^{+\beta h \p0}\eta_{\p0}^{-1}(x+\i)\eta_{\p0}^{-1}(x-\i)=1+\eta_{\p0-2}(x).
\eea
These equations can be brought into the form of $\p0$-many non-linear integral
equations identical to the TBA equations of \cite{Takahashi99} that are the
starting point of \cite{Zotos98}.\\

We want to present a method of computation of the relevant functions avoiding
an explicit solution of the TBA equations.  To this end we observe two
important properties of the functions $\eta_\alpha$:\\

(i) All objects $\Y_\alpha$ and $K$ are even functions of the magnetic field
$h$. Hence the first order derivatives of $\eta_\alpha$ with respect to $h$
evaluated at $h=0$ yield zero for all $\alpha$ except $\alpha=\p0-1$, $\p0$
where we have 
\be
\partial_{\beta h}\log\eta_{\p0-1}=\partial_{\beta
h}\log\eta_{\p0}=\p0/2.\label{derivwrth}
\ee

(ii) For vanishing magnetic field we have a remarkably simple identity
\bea
&&1+\eta_{\p0-1}(x)=1+\eta_{\p0}^{-1}(x)\cr
&&=\p0\frac{Q(x+\i(\p0-1)Q(x-\i(\p0-1)}
{\left[\sinh\left(\frac\gamma 2(x+\i (u+\p0-1))\right)
\sinh\left(\frac\gamma 2(x-\i (u+\p0-1))\right)\right]^{N/2}}\label{reduk}
\eea
involving the function $Q$ that can be computed without solving the $\p0$-many
TBA equations. The alternative computation is done on the basis of a set of
non-linear integral equations for {\it two} functions $a$ and $\bar a$ \cite{KlumTBA}
\bea
\log a(x)=-\beta{\epsilon_+(x)}
&+&\int_{-\infty}^\infty\kappa(x-y)\log(1+a(y))dy\cr
&-&\int_{-\infty}^\infty\kappa(x-y+2\i)\log(1+\bar a(y))dy\cr
\log\bar a(x)=-\beta{\epsilon_-(x)}
&+&\int_{-\infty}^\infty\kappa(x-y)\log(1+\bar a(y))dy\cr
&-&\int_{-\infty}^\infty\kappa(x-y-2\i)\log(1+a(y))dy,
\label{NLIE}
\eea
where the driving terms $\epsilon_\pm$ and integration kernels $\kappa$ 
are explicitly given by
\bea
\epsilon_\pm(x)&=&J\frac\pi 2\frac{\sin\gamma}\gamma e_0(x)
\pm\frac{\pi}{2(\pi-\gamma)}h,\quad
e_0(x)=\frac{1}{\cosh\frac\pi 2 x},
\\
\kappa(x)&=&\frac{1}{2\pi}\int_{-\infty}^\infty\frac
{\sinh(\frac{\pi}{\gamma}-2)k}
{2\cosh k\sinh\left(\frac\pi\gamma-1\right)k}\e^{ikx}dk~.
\label{kernel}
\eea
In the course of the derivation of these two non-linear integral equations,
explicit expressions for the function $\log Q(x)$ occur. By use of these
results we find for \refeq{reduk} straightforwardly 
\bea
&&1+\eta_{\p0-1}(x)=1+\eta_{\p0}^{-1}(x)\cr
&&=\p0\cdot\exp\left(
\int_{-\infty}^\infty[\omega(x-y-\i)\log(1+a(y))-\omega(x-y+\i)\log(1+\bar
  a(y))]dy\right)
\label{etafromq}
\qquad
\eea
where the function $\omega(...)$ is given by
\be
\omega(x)=\frac{1}{2\pi}\int_{-\infty}^\infty\frac
{1}{2\sinh\left(\frac\pi\gamma-1\right)k}\e^{ikx}dk~.
\ee
The $k$-integral may be regularized by either choosing a path along the real axis
avoiding $k=0$ in the upper or in the lower half plane: the term in brackets 
on the r.h.s. of \refeq{etafromq} containing 
convolution terms of $\omega$ with $a$ and $\bar a$ does not depend on the
choice of regularization as
the asymptotics of $a$ and $\bar a$ are identical.\\

The last equations are the main result of this section. The r.h.s. of
\refeq{final} can be calculated as the terms with $1\le \alpha\le\p0-2$ are
zero and those with $\alpha=\p0-1, \p0$ are obtained from \refeq{derivwrth}
and \refeq{etafromq}. Most importantly, in this formulation the number $\p0$,
that used to be the level of closure of the TBA equations, only enters via the
anisotropy parameter $\gamma$. Hence, we may regard the above formulation as
an analytic continuation to all anisotropies $0<\gamma<\pi$.

Finally, we like to mention that the largest eigenvalue may be calculated from 
\be 
-\beta
f=\frac{1}{4}\int_{-\infty}^\infty e_0(x) \log\left[(1+a(x))(1+\bar
a(x))\right] dx
\label{EqFE}
\ee
where an irrelevant constant, i.e. independent of $T$ and $h$ has been ignored. 

\subsection{Numerical results}

\begin{figure}[!htb]
\vspace{1cm}
\centering
\includegraphics[width=0.90\textwidth]{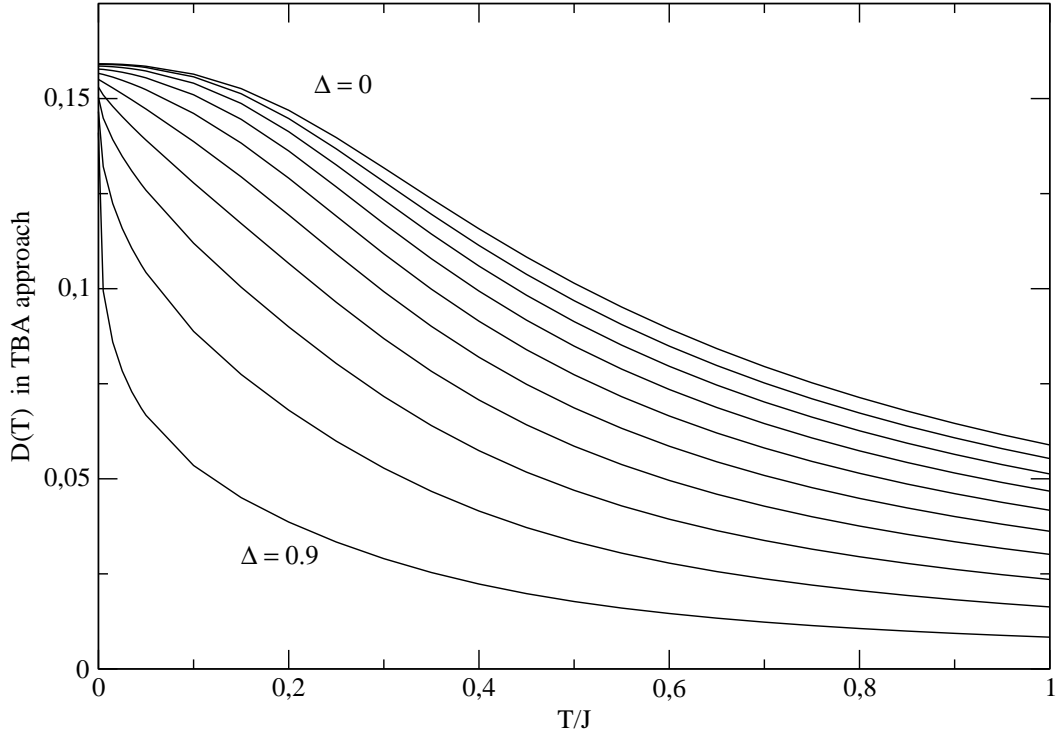}
\caption{The Drude weight in the temperature range $T/J=0,...,1$ for different
anisotropy parameters $\Delta=0,0.1,...,0.9$ as obtained in the TBA
approach. For $\Delta=1$ the corresponding data consist of a finite value of
the Drude weight for $T=0$ and zero for all $T>0$.}
\label{TBANum}
\end{figure}

Here we present the results of numerical treatments of the equations
(\ref{NLIE})-(\ref{etafromq}) derived above. For any anisotropy $0\le\Delta<
1$ we obtain Drude weights $D(T)$ that are monotonously decaying for
increasing temperature $T$, see Fig.\ref{TBANum}. In the zero temperature
limit the analytically known values $D_0$ are reproduced (the formula is given
in \refeq{D_0} below). At high temperatures the data $D(T)$ follow the
behaviour derived analytically in \refeq{DhighTBA}.

Most striking is the behaviour of $D(T)$ for parameters $\Delta$ close to
$1$. At $T=0$ the slope of $D(T)$ gets steeper the closer $\Delta$ approaches
$1$. For $\Delta=1$ the Drude weight shows singular behaviour with a finite value of
$D(T=0)$ and a drop to $D(T)=0$ for all $T>0$. This is in perfect agreement
with the results obtained in \cite{Zotos98} by a completely different and much
more elaborate treatment of the TBA equations comprising $\p0$ many non-linear
integral equations (with $\p0\to\infty$ for $\Delta\to 1$).

On the technical side we like to note that our treatment is not restricted to
anisotropy parameters $\Delta=\cos\frac\pi{\p0}$ with integer $\p0$. Also, we
conclude from the agreement of the two different numerical treatments of the
TBA equations (here and in \cite{Zotos98}) that both results are either
correct, or both are wrong in which case the failure must arise from assumptions
underlying the analytical derivation of \refeq{semifinal}. We will return to
this discussion at the end of the next subsection.\\

\subsection{High temperature asymptotics}
The analytical solution to the non-linear integral equations (NLIE)
\refeq{NLIE} in first order in $\beta=1/T$ is
obtained by writing the auxiliary functions
\begin{eqnarray}
\log{a}(x)&:=&-\beta e(x)+O(\beta^2)\nonumber\\
\log\overline{{a}}(x)&:=&-\beta\overline{e}(x)+O(\beta^2).
\end{eqnarray} 
These expressions are inserted into the NLIE and $\log(1+a), 
\log(1+\overline{a})$ are expanded up to first order in $\beta$ 
leading to the {\it linear} integral equations
\begin{eqnarray}
-\beta e(x)  &=&
-\beta\epsilon_+(x)\nonumber\\
&-&\frac{\beta}{2}\int\limits_{-\infty}^{\infty} \kappa\rundk{x-y}e(y)\,\,dy
+\frac{\beta}{2}\int\limits_{-\infty}^{\infty} \kappa\rundk{x-y+2\i}
\overline{e}(y)\,\,dy,\nonumber\\
-\beta\overline{e}(x)&=&-
\beta\epsilon_-(x)\nonumber\\
&-&\frac{\beta}{2} \int\limits_{-\infty}^{\infty} \kappa\rundk{x-y}\overline{e}(y)\,\,dy
+\frac{\beta}{2}\int\limits_{-\infty}^{\infty} \kappa\rundk{x-y-2\i}
e(y)\,\,dy.\nonumber\\
\end{eqnarray}
This system of linear integral equations is solved in Fourier space 
yielding explicit expressions for the Fourier transforms of 
$e$ and $\overline{e}$. Transforming back yields the result 
\begin{eqnarray}
\log{a}(x)&=&-\sin\gamma\,\beta 
i\rundk{\frac{2}{1-\e^{-\gamma x-\gamma\i}} -
\frac{1}{1-\e^{-\gamma x+\gamma\i}} -
\frac{1}{1-\e^{-\gamma x-3\gamma\i}}}-\beta h.\nonumber\\
\log\overline{{a}}(x)&=&-\sin\gamma\,\beta
i\rundk{\frac{1}{1-\e^{-\gamma x-\gamma\i}} +
\frac{1}{1-\e^{-\gamma x+3\gamma\i}}-
\frac{2}{1-\e^{-\gamma x+\gamma\i}} }+\beta h\nonumber\\
\end{eqnarray}
This has to be inserted into the expressions \refeq{etafromq} and
\refeq{final} for the Drude weight $D$. The integral was solved by means of
Mathematica yielding the
analytical high temperature result
\be
D\simeq \frac{C(\Delta)}T,\qquad C(\Delta)=J^2\frac{\gamma-\frac 12\sin 2\gamma}{16\gamma},
\label{DhighTBA}
\ee 
for $\gamma=\frac\pi\nu,\quad \nu=2, 3, ...$. This was compared to the
numerical data for $D(T)T$ obtained by numerical iteration of the NLIEs for
several anisotropy parameters in the high temperature limit. We observed very
good agreement in this regime with high accuracy. (Prior to publication of our
work we communicated the analytical results with the authors of \cite{ZP03}
who also found very good agreement with the numerical data of
\cite{Zotos98}.)\\

A fundamental problem appears with \refeq{DhighTBA} if we recall the unitary
transformation $U$ acting on every second lattice site $j$ by a rotation in
spin space by an angle $\pi$ around the $z$-axis and thus transforming $S^x_j,
S^y_j, S^z_j$ into $-S^x_j, -S^y_j, S^z_j$. Actually, in fermionic operator
language this is equivalent to changing $c_{j}^\dag, c_j$ to $-c_{j}^\dag,
-c_j$ on every second site without affecting the anticommutation rules. Hence
we have $U \hat{H}(J,\Delta) U^{-1}= \hat{H}(-J,-\Delta)$ and from
\refeq{curv} we conclude 
\be
D(T,J,\Delta)=D(T,-J,-\Delta)=-D(-T,J,-\Delta),\label{maptonegT}
\ee
where we also used the fact that $U$ does not change the dependence of
the energy levels on the twist angle $\phi$.\\

Equation \refeq{maptonegT} relates the Drude weight at positive temperature
$T$ and interaction parameters $J,\Delta$ to the Drude weight at {\it
negative} temperature $-T$ and interaction parameters $J,-\Delta$. In general
we do not learn much from this with respect to the properties of a system at
``physical'' temperatures, e.g. for $T\to 0+$ equation \refeq{maptonegT}
relates the Drude weight of the ground state of the system with $J$, $\Delta$
to the Drude weight of the highest energy state of the system with $J$,
$-\Delta$.  Quite differently, for high temperatures the asymptotical
behaviour \refeq{DhighTBA} is valid for large positive {\it and} negative
temperatures $T$! In fact, a high temperature series in $\beta\ (=1/T)$ is
meaningful and has to have some finite convergence radius such that
sufficiently small, but otherwise arbitrary (complex) arguments $\beta$ are
allowed.  Hence we may apply \refeq{DhighTBA} in the form $D\simeq
C(J,\Delta)/T$ for positive and negative temperatures and from
\refeq{maptonegT} we obtain $C(J,\Delta)=C(J,-\Delta)$ or
$C(\gamma)=C(\pi-\gamma)$ a relation certainly not satisfied by
\refeq{DhighTBA}.  There are several possible reasons for this failure. It may
simply be that \refeq{DhighTBA} is valid only for the discrete values of
$\gamma=\frac\pi\nu$ with $\nu=2, 3, ...$. In other words, the analytical
continuation is simply not allowed (at least not in the ``most natural''
way). We do not want to discuss this point any further.
\\

Another reason may be due to assumptions usually taken for granted in TBA
calculations, but failing in the derivation of \refeq{semifinal}. We already
pointed out that there are convergence problems in the transformation of
\refeq{d1} to \refeq{semifinal}. Yet another problem is due to the fact that
\refeq{semifinal} has been applied to the Heisenberg chain by dealing with
magnons and their bound states (``string'') as elementary, i.e. stable
particles. This is --as we nowadays know-- not strictly the case. The perfect
string picture may be strongly violated, especially if the string is located
at large spectral parameters.  This explanation is actually corroborated by
numerical studies \cite{GlocKlum}.

\section{TBA based on spinons and antispinons}

The above set of NLIEs \refeq{NLIE} has been derived in the QTM setting
\cite{KlumTBA} which is mathematically quite different from the TBA. In this
sense the QTM approach is an independent method based on algebraic and
analytical reasoning rather than combinatorial arguments in the case of the
TBA approach. However, the structure of the NLIEs \refeq{NLIE} is similar to
that of TBA. In fact, it has been argued \cite{KlJohn} that \refeq{NLIE} can
be viewed as the TBA equations of spinons and anti-spinons with bare energies
and scattering data given in \refeq{kernel} (with $\eta_1:=1/a$ and
$\eta_2:=1/\bar a$). Such a description of the thermodynamics of a system
based on its exact low energy excitations would usually be expected to be
possible, but restricted to the low temperature regime. Surprisingly, for the
thermodynamical potential of the Heisenberg chain it is quantitatively correct
for {\it all} temperatures and fields! Note also that \refeq{NLIE} is valid
for all $\gamma$, not just for the discrete set for which the number of bound
states of magnons is finite.\\

The idea of our work in this section is to utilize the spinon and anti-spinon
particle basis for describing the spectrum of the model. The advantage in
comparison to the basis of magnons and their bound states is obvious: from the
beginning there will be just two coupled NLIEs to be solved and only two terms
in the sum of \refeq{semifinal} in contrast to in general infinitly many for
generic values of $\Delta$. Also we no longer deal with disintegrating or
deformating strings. Of course, this approach is phenomenological and for the
computation of the Drude weight it can not be expected to be as successful as
for the free energy. With respect to the justification of our approach and its
limitation to low temperatures some arguments are given in the appendix.\\

The reader short of time may directly go to equation \refeq{final3} obtained
from \refeq{final} for just two particles: spinon and anti-spinon with
$\eta_1:=1/a$ and $\eta_2:=1/\bar a$. In the following treatment we want to
address certain subtle, but important issues like the response of spinons and
anti-spinons to twist angles $\phi$ and magnetic fields $h$. In section 2, the
two types of response were assumed for any particle species $\alpha$ to be
governed by just one number $n_\alpha$ which is usually an integer. From
\refeq{kernel} we see that the Zeeman energy for spinons involves a number
$\pm\frac{\pi}{2(\pi-\gamma)}$ that in general is non-integer. From
\cite{KBP91} we may infer that also the response to the twist angle $\phi$ is
governed by the same number. Here however, we want to treat the problem in a
different way by a mapping to a system where the numbers $n_\alpha$ are just 1
as usual. This system will involve just one {\it closed} integration contour
thereby also avoiding the divergence problem mentioned in the final paragraph
of the preceding section.\\

To achieve the outlined goals we reformulate the equations \refeq{NLIE} from
which we see that the functions $a(x)$ and $\bar a(x)$ can be analytically
continued into the complex plane where they turn into each other due to the
identity $\bar a(x)=1/a(x-2\i)$.\\

Substituting $\log(1+a)=\log(1+1/a)+\log a$ 
in \refeq{NLIE} and resolving for $\log a$ yields
\bea
\log\eta_+(x)=\beta{\epsilon(x)}
&-&\int_{-\infty}^\infty\varkappa(x-y)\log(1+\eta_+^{-1}(y))dy\cr
&+&\int_{-\infty}^\infty\varkappa(x-y+2\i)\log(1+\eta_-^{-1}(y))dy,
\label{NLIE2}
\eea where $\eta_+(x):=a(x)$ and $\eta_-(x):=1/\bar a(x)$. The function
$\epsilon(x)$ happens to be the energy of a single magnon parametrized
by the spectral parameter $x$
\bea
\epsilon(x)&=&-J\frac{\sin\gamma}\gamma\, p'(x)-h,\cr
p(x)&=&-i\log\frac{\sinh(\frac\gamma 2 (x-\i))}{\sinh(\frac\gamma 2(x+\i))},
\eea
and $p(x)$ is the momentum. Furthermore, $\varkappa(x)$ is given by
\be
\varkappa(x)=\frac{1}{2\pi}\Theta'(x),\quad
\Theta(x):=-i\log\frac{\sinh(\frac\gamma 2 (x-2\i))}{\sinh(\frac\gamma 2 (x+2\i))},
\ee
where $\Theta(x-y)$ is the scattering phase of two magnons with spectral 
parameters $x$ and $y$.

Due to the relation $\eta_-(x+2\i)=\eta_+(x)$ we can immediately write down
a second equation from \refeq{NLIE2} yielding a complete
set of equations for $\eta_\pm$.\\

These equations are quite different from the standard TBA equations
for the spin-1/2 Heisenberg model since those are based on up to infinitely
many bound states (strings). However, the above equations may be obtained
within the TBA method if the following two modifications with respect to
the standard approach are applied
\begin{itemize}
\item ignore bound states,

\item consider the single magnon with energy $\epsilon(x)$ and momentum $p(x)$
parametrized by the spectral parameter $x$ on the real axis Im $x=0$ 
(``$C_+$'') as well as on the axis Im $x=-2$ (``$C_-$'').
\end{itemize}

{\it A posteriori} we find that this prescription renders the single
magnon a complete particle basis for the thermodynamics of
the system. In other words, if we take the magnon with bare momentum $p$ 
and energy $\epsilon$ on the axes $C_\pm$ 
the TBA method of the previous section yields exactly \refeq{NLIE2}. The
free energy then reads
\begin{equation} 
-\beta f=\frac{1}{2\pi}\sum_{\alpha=\pm}\int_{-\infty}^{\infty}
p_\alpha'(x)\ln(1+\eta^{-1}_\alpha(x))\,\,dx.
\label{FreeEn}
\end{equation}
where  $p_+(x)=p(x)$, $p_-(x)=-p(x-2\i)$ (both functions are monotonously
increasing with increasing $x$) and
$\eta_\alpha$ are determined from \refeq{NLIE2} or
\begin{equation}
\ln \eta_\alpha(x)=\beta\epsilon_\alpha(x)
-\frac{1}{2\pi} \sum_\beta \KK_{\alpha \beta}\ast \ln\rundk{
1+{\eta_\beta}^{-1}},
\label{rho2a}
\end{equation}
with  $\epsilon_+(x)=\epsilon(x)$, $\epsilon_-(x)=\epsilon(x-2\i)$ and
$\KK_{++}(z)=\KK_{--}(z)=\KK(z)$, 
$\KK_{+-}(z)=-\KK(z+2\i)$, $\KK_{-+}(z)=-\KK(z-2\i)$.\\

Finally, we note for the Drude weight formula \refeq{final} with $\alpha$
ranging only over $\alpha=+,-$. As \refeq{bareen} has to be slightly modified 
by $\pm$ signs, because of $\epsilon_\pm=\pm J\frac{\sin\gamma}\gamma\,p_\alpha'$, we find
\refeq{diffid} consequently modified to 
$J\frac{\sin\gamma}\gamma\, \rho_\alpha=\mp\frac{1}{2\pi}\frac{\partial}{\partial\beta}
\ln(1+\eta_\alpha^{-1})$. The Drude weight is
\be
D=\frac{J\sin\gamma}{4\pi\beta\gamma}\sum\limits_{\alpha=\pm}\alpha
\int_{-\infty}^\infty\frac
{(\frac{\partial}{\partial\beta h} \ln\eta_\alpha)^2
(\frac{\partial}{\partial\lax} \ln\eta_\alpha)^2}
{(1+\eta_\alpha)(1+\eta_\alpha^{-1})
\frac{\partial}{\partial\beta}\ln\eta_\alpha}
d\lax.
\label{final2}
\ee
Alternatively, in terms of
$a $ $(=\eta_+)$, $\bar a $ $(=\eta_-^{-1})$ and due to the relation
\refeq{particlehole} we get
\be
D=\frac{J\sin\gamma}{4\pi\beta\gamma}\sum\limits_{\alpha=1, 2}
\int_{-\infty}^\infty\frac
{(\frac{\partial}{\partial\beta h} \ln a_\alpha)^2
(\frac{\partial}{\partial\lax} \ln a_\alpha)^2}
{(1+a_\alpha)(1+a_\alpha^{-1})
\frac{\partial}{\partial\beta}\ln a_\alpha}
d\lax.
\label{final3}
\ee 
with the notation $a_1:=a$, $a_2:=\bar a$. This expression with functions $a$
and $\bar a$ calculated from \refeq{NLIE} is our final analytic result for the
finite temperature Drude weight within the spinon approach. In the next
subsections we will numerically evaluate $D$ for arbitrary temperatures and
study analytically the zero temperature limit and the high temperature asymptotics.

\subsection{Numerical results}
\begin{figure}[!htb]
\vspace{1cm}
\centering
\includegraphics[width=0.90\textwidth]{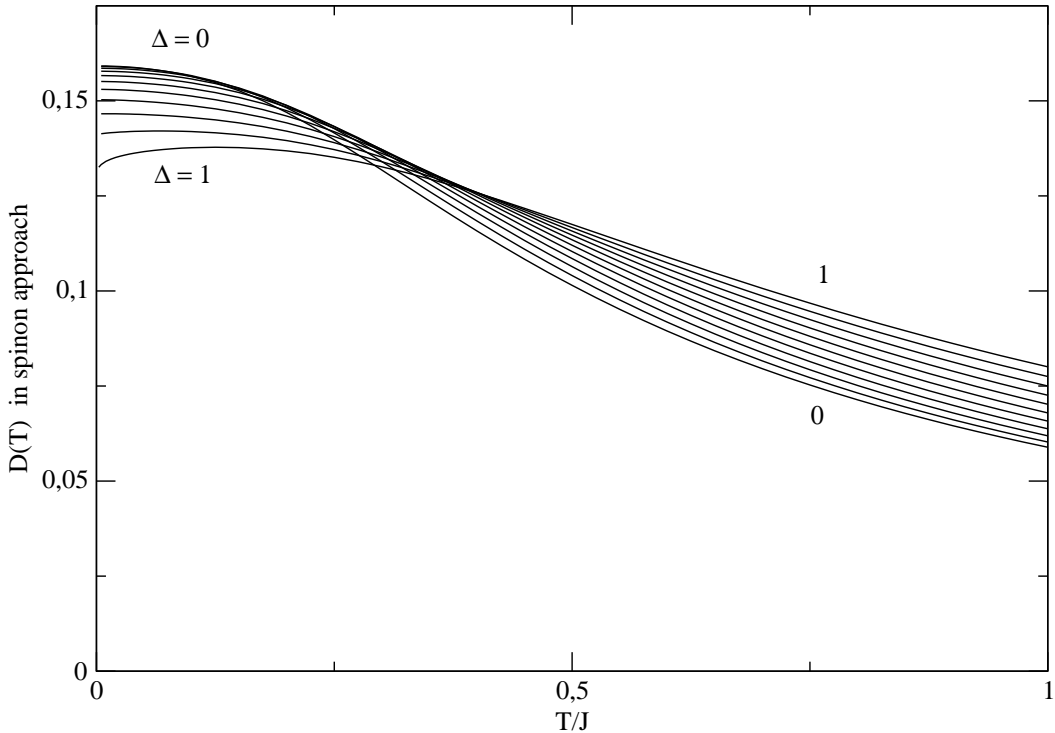}
\caption{The Drude weight in the temperature range $T/J=0,...,1$ for different
anisotropy parameters $\Delta=0,0.1,...,0.9, 1$ as obtained in the spinon
approach.}
\label{SpinonNum}
\end{figure}

Our numerical results are obtained for
arbitrary temperatures $T$ and various anisotropy parameters
$\Delta=\cos\gamma$ in the repulsive regime $[0,1]$,
cf. Fig.\ref{SpinonNum}. We find qualitatively different behaviour
depending on the value of $\Delta$.\\

\begin{figure}[!htb]
\vspace{1cm}
\centering
\includegraphics[width=0.90\textwidth]{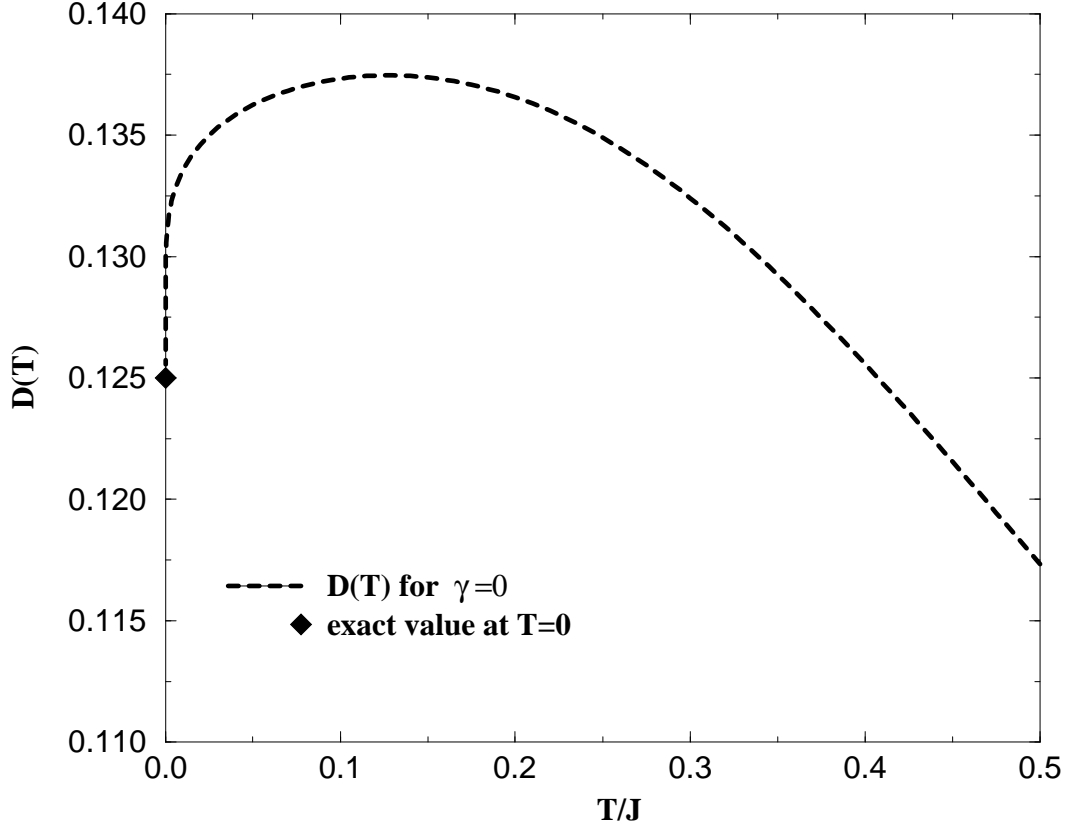}
\caption{The Drude weight at the isotropic point $\Delta=1$ ($\gamma=0$). A very steep
slope close to $T=0$ is observed. This slope is infinite at precisely $T=0$.}
\label{log}
\end{figure}

For $\Delta$ close to the free fermion point $\Delta=0$ the Drude weight
$D(T)$ is a monotonously decreasing function of temperature, see
Fig.\ref{SpinonNum}. However, for larger values of $\Delta$ close to
$\Delta=1$ corresponding to the isotropic antiferromagnetic Heisenberg point
the dependence of $D(T)$ on temperature is non-monotonous! For sufficiently
low $T$ the function $D(T)$ {\it increases} with temperature.  After taking a
finite temperature maximum the function $D(T)$ decreases.  In particular, the
values of $D(T)$ at the isotropic point ($\Delta=1$) are non-zero for any
value of temperature $T$, see Fig.\ref{log}. This is in striking contrast to
the results of \cite{Zotos98} and those of the preceding section. A discussion
of this will be given in the final section of this paper.\\

\subsection{Drude weight at $T=0$}
In the limit of $T\to 0$ and large argument $x$ 
we have the scaling behaviour of the driving term in the NLIE
\begin{equation}
\frac{\beta}{\cosh\frac{\pi}{2}x}\rightarrow 2\beta
\e^{-\frac{\pi}{2}\abs{x}}.
\end{equation}
From this observation it was concluded \cite{KlumTBA} that the leading low $T$
asymptotics of the physical properties is determined by values of the spectral
parameter $x\simeq\pm\frac 2\pi\ln\beta$.  Indeed we find that
$\lim_{T\to0}a_\alpha(x\pm\frac 2\pi\ln\beta)$ yields a well defined
function of $x$ still satisfying a non-linear integral equation. However the
associated functions like $\frac{\partial}{\partial x}a_\alpha$ and
$\frac{\partial}{\partial\beta}a_\alpha$ satisfy {\it linear} integral
equations. Actually, both sets of integral equations are identical up to
different driving terms. Those, however, are strictly proportional to
each other as they are given by
\begin{eqnarray} 
\delieins{x}\rundk{\frac{\beta}{\cosh\frac{\pi}{2}x}}
&\rightarrow&\mp\pi\beta\e^{-\frac{\pi}{2}\abs{x}},\nonumber\\
\delieins{\beta}\rundk{\frac{\beta}{\cosh\frac{\pi}{2}x}}
&\rightarrow&2\e^{-\frac{\pi}{2}\abs{x}},\nonumber\\
\end{eqnarray} 
with proportionality factor $\pm\frac\pi 2\beta$. Therefore, in the
limit $T\rightarrow 0$ and for the relevant range of spectral parameters
$x$ the derivatives of $\ln a_\alpha$ with respect to $x$ and $\beta$ satisfy
\begin{equation}
\frac{\delieins{x}\ln a_\alpha}{\delieins{\beta}\ln a_\alpha}
=\mp\frac{\pi}{2}\beta.
\end{equation}
Hence in \refeq{final3} we find the simplification
\be
D_0=\frac{J\sin\gamma}{4\gamma}\sum\limits_{\alpha=1, 2}
\int_{0}^\infty\frac
{(\frac{\partial}{\partial\beta h} \ln a_\alpha)^2
(\frac{\partial}{\partial\lax} \ln a_\alpha)} {(1+a_\alpha)(1+a_\alpha^{-1})}
d\lax.
\ee
By use of the ``dressed function'' formalism we obtain the identity
\be
\sum\limits_{\alpha=1, 2}\int_{0}^\infty
\frac{(\frac{\partial}{\partial\beta h}a_\alpha)^2
\frac{\partial}{\partial\lax}a_\alpha}{a_\alpha^2(1+a_\alpha)^2}dx
=-\beta
\sum\limits_{\alpha=1, 2}\int_{0}^\infty
\frac{\partial^2}{\partial(\beta h)^2}[\ln(1+a_\alpha)]
\frac{\partial}{\partial\lax}\epsilon_\alpha dx,
\ee
Hence the Drude weight is
\be
D_0=\frac{\beta}{4}\left(\frac\pi 2\frac{\sin\gamma}{\gamma}J\right)^2
\frac{\partial^2}{\partial(\beta h)^2}\sum\limits_{\alpha=1, 2}
\int_{0}^\infty[\ln(1+a_\alpha)]e_0 dx,
\ee
where the term to the right of the partial derivative is identical to
$-2\beta f$. The second derivative with respect to $\beta h$ leads to the 
magnetic susceptibility 
\be
\chi_0=\frac 1{2v(\pi-\gamma)},\quad v=\frac\pi 2\frac{\sin\gamma}\gamma J.
\ee
Finally the zero temperature Drude weight is found to be
\begin{equation}
D_0=\frac{\pi\sin\gamma}{8\gamma(\pi-\gamma)}J.\label{D_0}
\end{equation}
This result agrees exactly with the Drude weight directly obtained from the
groundstate energy of a finite system with twist \cite{ShasS90,Zvyagin}.

\subsection{High temperature asymptotics}
\begin{figure}
\centering
\includegraphics[width=0.90\textwidth]{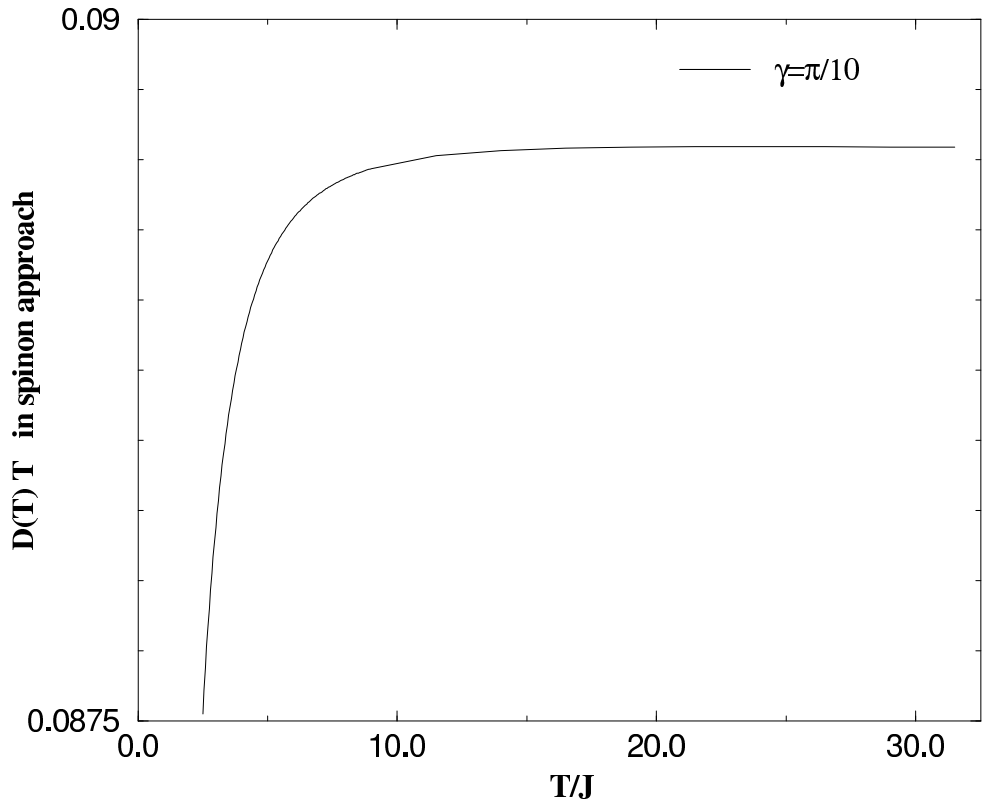}
\caption{The Drude weight at high temperature showing a $1/T$-decrease 
(results shown for $\gamma=\frac{\pi}{10}$).}
\label{hochT}
\end{figure}

As an illustration of the high temperature behaviour we show a plot of $T
D(T)$ for some value of $\Delta$ close to 1, see Fig.\ref{hochT}.
The analytic evaluation using \refeq{DhighTBA} in \refeq{final3} yields
\be
D\simeq \frac{C(\Delta)}T,\qquad C(\Delta)=J^2\frac{\Delta^2+2}{32}
\label{Cinspinon}
\ee
Again there is a problem with the analytically derived expression for $C(\Delta)$.
Here it contradicts certain rigorously known properties of the high temperature
limit that we are going to derive from the spectral representation
\refeq{spectralD}. First we symmetrize the summation
\be
D=\frac 1L\Bigg[\frac 12\underbrace{\langle -\hat{T}\rangle}_{\displaystyle{\ge 0}} \quad
-\underbrace{\sum_{m\not=n}p_n\frac{|\langle n|J|m\rangle|^2}
{\epsilon_m-\epsilon_n}}_{\displaystyle{=\frac 12\sum_{m\not=n}
\underbrace{\frac{p_n-p_m}{\epsilon_m-\epsilon_n}}_{\displaystyle{\ge 0}}
|\langle n|J|m\rangle|^2}}
\Bigg]\label{symmetrized}
\ee
As $D$ is the difference of two non-negative terms we see
\be
D\le\frac 1{2L}\langle -\hat{T}\rangle\label{UnglD}
\ee
where equality holds if the second term in \refeq{symmetrized} disappears;
this happens for $\Delta=0$ where the matrix elements $\langle n|J|m\rangle$
for different states $n$ and $m$ yield zero.

From direct calculations at high $T$ we find $\langle
-\hat{T}\rangle\simeq\widetilde C/T$ with a constant $\widetilde C$
independent of $\Delta$. From \refeq{UnglD} we see $C(\Delta)\le \widetilde
C/2L$. On the other hand we know that equality holds in the case $\Delta=0$,
therefore
\be 
C(\Delta)\le C(0),\label{inequality}
\ee
an inequality that has been obtained earlier on grounds of the optical sum
rule, see e.g. \cite{ZP03}.  This is clearly violated by our
analytic result \refeq{Cinspinon}!\\

The status of the high temperature results
in the spinon approach and those in the TBA approach are equally problematic:
in the spinon approach the inequality \refeq{inequality} is violated,
in the TBA approach the symmetry with respect to $\Delta$ is violated.

\section{Discussion}

In section 2 we reviewed the analytical method of computation of the
finite temperature Drude weight as proposed in \cite{FujKa98} and used in
\cite{Zotos98} for the study of the spin-1/2 Heisenberg chain. We then
employed this method as the starting point of our own analysis of the
Heisenberg chain. This was done in two different ways.\\

In section 3 we utilized the particle basis of magnons and their bound
states. Within this approach we managed to reduce the resultant equations to
only two non-linear integral equations. These results are totally equivalent
to those of \cite{Zotos98}, however, they allowed for an analytic continuation
to the treatment of all anisotropies $-1<\Delta\le 1$. Also, we derived an
analytic formula for the high temperature asymptotics.\\

In section 4 we employed the spinon and anti-spinon particle basis. The
results obtained in this approach strongly deviate from those of the preceding
section and hence from \cite{Zotos98}. We observed qualitatively different
behaviour of $D(T)$ for $\Delta$ close to 0 and 1 showing monotonous and
non-monotonous temperature dependence, respectively.  Instead of a drop of
$D(T)$ at $T=0$ our results show an increase with apparently infinite slope at
$T=0$, see Fig.\ref{log}. This is reminiscent of the behaviour of the magnetic
susceptibility with infinite slope due to logarithmic corrections at $T=0$.\\

Though disagreeing with \cite{Zotos98} we find strong similarities of our
results in the spinon approach with the numerical work \cite{NarMA98} which is
based on complete diagonalization of quantum chains up to length $L=14$.
Unfortunately, the non-monotonous dependence of the Drude weight data obtained
in \cite{NarMA98} may still be considered as plagued by finite size
corrections. In a later numerical analysis in \cite{MHCB2} the size dependence
was carefully studied for system sizes up to $N=18$. It was argued
\cite{MHCB2} that a non-monotonous behaviour found at $\Delta=1$ was not an
artefact of finite size effects and qualitatively agreed with our (then
unpublished) data that we presented here in section 4. Also, by use of quantum
Monte Carlo calculations at low temperatures, the authors of \cite{AlGros1}
found quantitative agreement of their results for anisotropy
$\Delta=\cos\pi/6$ with our (then unpublished) findings in the spinon approach,
which are in strong disagreement with \cite{Zotos98}.\\

Still, the numerical treatments have to be considered with care as the size
dependence is indeed very strong. However, also the analytical treatments have
to be considered with care. For the case of the TBA approach on the basis of
bound states (strings) we have indicated certain problems and mechanisms of
possible failure. In our understanding, signatures of this are even visible at
high temperatures, cf. subsections 3.2 and 4.3. This will be studied in more
detail in \cite{GlocKlum}.\\

In the case of the spinon approach we are confident that the results are reliable
at low temperature, whereas they are not trustworthy at high temperatures. The
reason for this is simply that the concept of the spinon and anti-spinon
particles is field theoretic and restricted to low energies and low
temperatures. (Strangely, for the static properties encoded in the free energy,
the spinon-concept gives correct results for {\it all} temperatures and
fields! Some further aspects along this line are discussed in the appendix.)\\

Finally, we like to mention that our results in the spinon approach have been
confirmed by conformal field theoretical arguments developed in \cite{FK03}
for certain anisotropy values $\Delta$, however for some other values (notably those
close to 1) there is still strong disagreement. A comparison of all data 
in collaboration with the authors of \cite{FK03} is on the way.

\section*{Acknowledgments}
The authors like to acknowledge valuable discussions with H. Frahm, S. Glocke,
S.~Fujimoto, F. Heidrich-Meisner, N. Kawakami, A. Rosch, K. Sakai and in particular
X. Zotos.  The hospitality of Cologne University and Dormund University is
appreciated where large parts of the work were performed.  A.K. acknowledges
financial support by the {\it Deutsche Forschungsgemeinschaft} under grant No.
Kl~645/3-3, 4-1, 4-2, and encouragement by {\it Schwerpunkt SP1073}.

\renewcommand{\theequation}{A.\arabic{equation}}

\section*{Appendix}

\setcounter{equation}{0}

Here we address the question whether the relations \refeq{NLIE2} and
\refeq{FreeEn} are more than accidental. We want to argue that the
functions $\rho_\alpha$ ($\alpha=\pm$) obtained from equations
\refeq{diffid} in the spinon approach (and satisfying (\ref{rho1},\ref{rho2}))
are true density functions. A (necessary) criterion to be satisfied
is the requirement that $\rho_\alpha$ determine all thermal expectation
values of the higher conserved quantities $F_n$ where 
\be
F_n=\frac{\partial^n}{\partial x^n}\log T(x)\Big|_{x=0}, 
\ee 
and $T(x)$ is the
six-vertex model row-to-row transfer matrix in dependence on the
spectral parameter $x$ with decoupling point $x=0$. For $n=0, 1$ we have
$F_0\sim P$ (momentum operator) and $F_1\sim H$ (Hamiltonian). 
We are interested
in the cases $n\ge 1$ and define the differential operator $D_n={\partial^n}/
{\partial x^n}$.\\

The contribution of a magnon with spectral parameter $x$ to the eigenvalue
of any $F_n$ is
$(D_{n}p)(x)$. If $\rho_\alpha$
determine all thermal measurements we must find the relation 
\be
\langle F_n\rangle=
\sum_{\alpha=\pm}\int_{-\infty}^{\infty}
(D_{n}p_\alpha)(x)\rho_\alpha(x)\,\,dx.
\label{Fnavexp}
\ee

We want to verify this expectation and calculate
\be
\langle F_n\rangle=\langle D_n\log T(x)\rangle=
\frac{\partial}{\partial z}\log\Tr \left[\exp(1+z D_n\log T)
\e^{-\beta H}\right],
\label{Fnav}
\ee
where the derivative is to be taken at $z=0$. The reason for introducing
the rather involved expression on the r.h.s. is that exactly this quantity
can be calculated in the QTM approach resulting into almost literally 
\refeq{rho2a} except the replacement
\be
\beta\epsilon_\alpha(x)\to(\beta-zD_{n-1})\epsilon_\alpha(x),
\label{replace}
\ee
which can be derived by simple but lengthy calculations.
Note the subscript $n-1$! The free energy (or more precisely the quantity
$\log\Tr[...]$ in \refeq{Fnav}) is still given by the r.h.s. of \refeq{FreeEn}. Hence
the thermal average of $F_n$ is
\begin{equation} 
\langle F_n\rangle=\sum_{\alpha=\pm}\int_{-\infty}^{\infty}
\frac{p_\alpha'(x)}{2\pi}\frac{\partial}{\partial z}
\ln(1+\eta^{-1}_\alpha(x))\,\,dx.
\label{FreeEn1}
\end{equation}

The expressions \refeq{Fnavexp} and \refeq{FreeEn1} look quite
different.  Still they are equivalent as can be seen in the ``dressed
functions'' formalism. To this end we note \refeq{rho2a} with the replacement
\refeq{replace}, and take the derivative with respect to $z$ at $z=0$
\begin{equation}
(1+\eta_\alpha)\frac{\partial}{\partial z}\ln\rundk{1+{\eta_\beta}^{-1}}=
-D_{n-1}\epsilon_\alpha
+\frac{1}{2\pi} \sum_\beta \KK_{\alpha \beta}\ast 
\frac{\partial}{\partial z}\ln\rundk{1+{\eta_\beta}^{-1}}.
\label{rho3}
\end{equation}
This equation can be regarded as an integral equation for
$\frac{\partial}{\partial z}\ln\rundk{1+{\eta_\beta}^{-1}}$ and is
very similar to that one for $\rho_\alpha$ \refeq{rho1}, just the
inhomogeneity (driving term) is different
\begin{equation}
(1+\eta_\alpha)\rho_\alpha (\lax)=
\frac{1}{2\pi } p'_\alpha(\lax) + \frac{1}{2\pi}\sum_\beta 
\KK_{\alpha \beta }\ast \rho _\beta.
\label{rho4}
\end{equation}
It is now a standard exercise in mathematical analysis to show the identity
of \refeq{Fnavexp} and \refeq{FreeEn1}.\\

Unfortunately, the criterion that the functions $\rho_\alpha$ yield the
correct expectation values is necessary, but not sufficient. True density
functions are usually real and non-negative. In our case, however, complex valued
functions are in principle allowed as we may have distributions along curved
lines in the complex plane. Along these lines just the ratio of particle and
hole densities (i.e. the function $\eta_\alpha$) must take positive values. In
the spinon approach this is satisfied for low temperatures, but not for high
temperatures.

\end{document}